\documentclass[useAMS,usenatbib]{mn2e}
\usepackage{graphicx}
\usepackage{txfonts}

\title[Differential stellar population models]{Differential stellar population models: how to reliably measure [Fe/H] and [$\alpha$/Fe] in galaxies}

\author[Walcher, Coelho, Gallazzi \& Charlot]{
C.J. Walcher$^{1}$\thanks{E-mail:jwalcher@rssd.esa.int}, P. Coelho$^{2}$, A. Gallazzi$^{3}$, and S. Charlot$^{2}$\\
$^{1}$Research and Scientific Support Department, European Space Agency, Keplerlaan 1, 2200AG Noordwijk, The Netherlands\\
$^{2}$Institut d'Astrophysique de Paris, CNRS, Universit\'e Pierre \& Marie Curie, UMR 7095, 98bis, bd Arago, 75014 Paris, France \\
$^{3}$Max Planck Institut f\"ur Astronomie, K\"onigstuhl 17, 69117 Heidelberg, Germany
}
\begin{document}

\date{}

\maketitle

\begin{abstract}

We present Ódifferential stellar population modelsÓ, which allow improved determinations of the
ages, iron and $\alpha$-element abundances of old stellar populations from spectral fitting. These
new models are calibrated at solar abundances using the predictions from classical, semi-empirical
stellar  population models. We then use the predictive power of fully synthetic models to compute
predictions for different [Fe/H] and [$\alpha$/Fe]. We show that these new differential models provide
remarkably accurate fits to the integrated optical spectra of the bulge globular clusters NGC6528
and NGC6553, and that the inferred [Fe/H] and [$\alpha$/Fe] agree with values derived elsewhere from stellar
photometry and spectroscopy. The analysis of a small sample of SDSS early-type galaxies further
confirms that our $\alpha$-enhanced models provide a better fit to the spectra of massive ellipticals 
than the solar-scaled ones. Our approach opens new
opportunities for precision measurements of abundance ratios in galaxies.

\end{abstract}


\section{Introduction}

Stellar population models designed to understand the photospheric emission from 
stars in galaxies and star clusters have reached an impressive degree of maturity 
in the last decade \citep[e.g.][]{vazdekis99,bruzual03,le-borgne04,vazquez05}. 

A remaining problem is that theoretical stellar spectra are not 
mature enough to reproduce real stars at all spectral types and complete wavelength 
range in an absolute sense \citep[e.g.][]{martins07,bertone08}. The most common 
solution is to replace in the optical wavelength regime the theoretical stellar spectra 
with spectra derived from empirical spectral libraries. We call such models "semi-empirical". 
A draw-back of these models has always been that the 
stellar spectra are restricted to the abundance patterns present in the empirical libraries 
(dominated by solar neighbourhood stars). So, while semi-empirical models provide an excellent basis 
for solar metallicity and [$\alpha$/Fe]=0 \citep[e.g.][]{asari07,koleva08}, they are 
biased at non-solar metallicites. 

To overcome these biases, we here use the predictive power of the fully theoretical 
models from \citet[][hereafter C07]{coelho07} to compute \emph{relative} changes 
in flux F$_{\lambda}$
as a function of [Fe/H] and [$\alpha$/Fe]. These are then applied to the solar 
abundance SSPs from semi-empirical models, yielding "differential population 
synthesis models" that are at the same time highly accurate in an absolute sense 
and whose predictions for non-solar abundance patterns are well-defined and 
unbiased. The new differential models are accurate enough to 
reliably determine the (over-)abundance of the $\alpha$-elements 
from the integrated spectrum of an 
old stellar population. Predicting and fitting every pixel in the spectrum, 
as compared to predicting the Lick indices "only" \citep{trager00,thomas04}, 
allows one to (1) use the redundancy of the information at different 
wavelengths, for data of lower S/N, (2) analyse spectra at higher resolution 
(3) explore wavelength ranges not covered by Lick indices and (4) free oneself 
of potentially complicated corrections for instrumental and velocity dispersion.

\section{Computing differential models}

Semi-empirical stellar population models provide a reliable zero-point for the spectral 
evolution of stellar populations of solar metallicity (more precisely [Fe/H]=0 and [$\alpha$/Fe]=0, 
called "solar" for the remainder of this letter). We therefore rely on them for the absolute 
calibration (including their stellar evolutionary tracks). 
For the present letter, we used \citet[][hereafter BC03]{bruzual03}, 
\citet[][hereafter PegHR]{le-borgne03}, and an updated version of 
\citet[][hereafter V09]{vazdekis99} and we refer to these papers for a complete 
description. 

For the pixel-by-pixel corrections (subsequently 
called "differentials") as a function of [Fe/H] and [$\alpha$/Fe] we rely on the purely theoretical 
models computed by C07 and we refer to this paper for all details. 
The models predict the spectra of SSPs in the age range 3 to 14 Gyr, in the wavelength 
range 3000 {\AA} to 1.34 $\mu$m, and for the abundance values [Fe/H] = $-0.5$, 0.0 and 
0.2  and [$\alpha$/Fe] = 0.0 and 0.4 at a constant resolution of full width at half-maximum 
(FWHM) = 1 {\AA}. We additionally interpolated the models for [Fe/H] = $-0.25$ and 
[$\alpha$/Fe] = 0.2 in order to obtain a finer grid. The interpolation is done 
linearly in log(Flux). This is because the absorption caused by a single line of an  
element species is proportional to the exponentiated opacity, while the opacity itself 
depends linearly on element abundance. We will refer to this grid of 120 SSPs as 
the C07 models. We make the usual assumption that the fully theoretical models 
correctly predict the flux variation as a function of varying [Fe/H] and [$\alpha$/Fe] 
\citep[e.g.][]{trager00,thomas04}.  

The differential recalibration procedure involves the following steps: (1) The SSP spectrum by
C07 is convolved with a gaussian filter and then resampled to achieve the same resolution 
and binning as the semi-empirical SSP spectra used for calibration. (2) At a given age, 
the differential at each pixel as a function of [Fe/H] and at fixed [$\alpha$/Fe]=0 
is defined as the flux ratio of the SSP 
flux of a given [Fe/H] to the spectrum with [Fe/H] = 0.0. (3) Similarly, the ratios between the 
$\alpha$-enhanced fluxes and solar-scaled fluxes, at the same [Fe/H], give the differential 
as a function of [$\alpha$/Fe]. (4) All semi-empirical models were interpolated in log(Flux) 
to provide the same SSP ages as C07. (5) The last step is then to multiply the solar SSP 
fluxes of the semi-empirical models by the correction vectors derived from the fully theoretical 
models. 

As a result we obtain three sets of differential models (one for each calibrator)
at the same grid points in age, [Fe/H] and [$\alpha$/Fe] as the C07 models.
We will refer to these as BC03+C07, PegHR+C07, and V09+C07, 
respectively. 

\begin{figure}
\begin{center}
\resizebox{0.90\hsize}{!}{\includegraphics[]{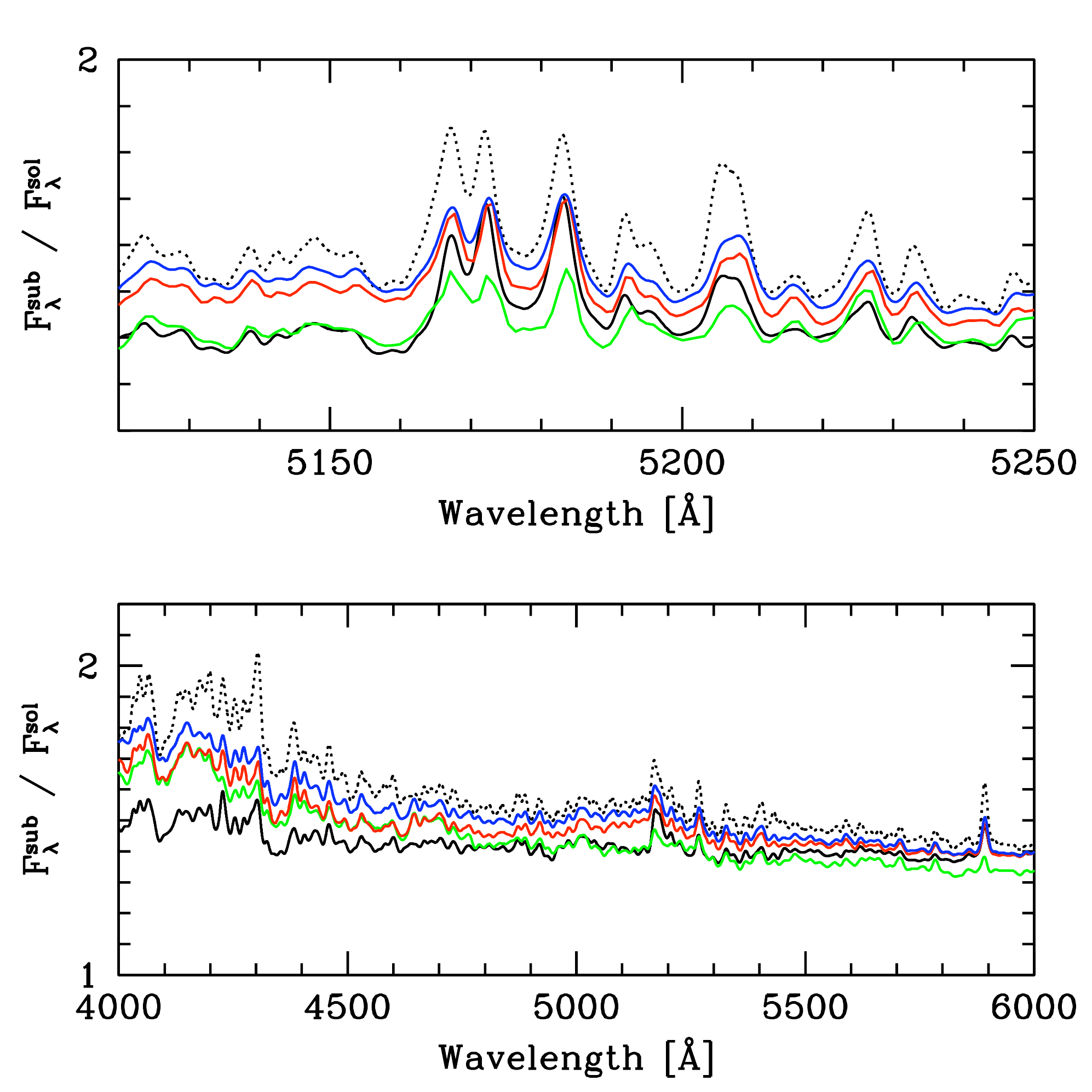}}
\caption[differentials]{The ratios between SSPs of 10Gyr at subsolar metallicity (Z=0.005, [Fe/H]=$-0.5$, 
[$\alpha$/Fe]=0) and solar metallicity ([Fe/H]=0,[$\alpha$/Fe]=0)) for different 
stellar population models (black C07, green BC03, blue PegHR and red V09). The dotted 
line is C07 at the same Z=0.005, but different abundance ratios ([Fe/H]=$-0.85$ and [$\alpha$/Fe]=0.4).
All models have been smoothed to a common resolution of $\sigma$=1{\AA} in the upper panel and 
$\sigma$=5{\AA} in the lower panel. }
\label{f:diff}
\end{center}
\end{figure}

For illustration we show one of the derived differentials, namely for [Fe/H]=$-0.5$ 
and [$\alpha$/Fe]=0.0 (Z=0.005), in Figure \ref{f:diff}. It is clear that the main signal 
of the variation in [Fe/H] is constrained to specific wavelength ranges, such as e.g. 
the MgB feature. For comparison with previous work we also show the differentials 
of the semi-empirical models in Figure \ref{f:diff}. We interpolated the semi-empirical 
models in log(Flux) to fall on [Fe/H]=$-0.5$\footnote{The models are generally defined 
in total metallicity Z and not in [Fe/H]; under the assumption of solar abundance ratios, i.e. 
[$\alpha$/Fe]=0, [Fe/H]=$-0.5$ is equivalent to Z=0.005.}.

The differentials derived from the semi-empirical SSPs are significantly bluer 
than their theoretical counterparts. Insight into this problem can be gained by 
overplotting in Figure \ref{f:diff} the differential of the C07 model for the 
same total metallicity (Z=0.005) but with [$\alpha$/Fe]=0.4 and therefore 
[Fe/H]=$-0.85$ (obtained by extrapolation in log(Flux)). This new differential 
shows the same effect as the semi-empirical models, only even stronger. 
Indeed, the stars with sub-solar [Fe/H] represented in the stellar libraries 
are generally $\alpha$-enhanced due to the correlation between [Fe/H] 
and [$\alpha$/Fe] that is inherent to the solar neighbourhood 
\citep[see e.g. Figure 8 from][]{bensby05}. Semi-empirical models thus are 
biased to [$\alpha$/Fe]$>$0 at subsolar metallicities. The new differential 
models are free of this bias and even predict the change of the SSP spectra 
with [$\alpha$/Fe]. 

We emphasize that the above differential procedure accounts for the effects of
abundance variations both on stellar evolution and on stellar spectra.
It is therefore more accurate than methods based on the correction of
only stellar spectra \citep{prugniel07, cervantes07}. We note that the
difference in the stellar evolution prescriptions of the C07 and the semi-empirical
models should have only a minor influence on the differentials, as illustrated by
the similarity of the results obtained for semi-empirical models with 3 different
prescriptions in Figure \ref{f:diff}.


\section{Pixel-fitting of spectra}
\label{s:sedfit}

In order to test the applicability of the model spectra on observational 
data, we use a pixel-fitting software previously used and fully described in 
\citet{walcher06}, hereafter {\tt sedfit}. 
In brief, the software uses an initial template to obtain a reasonable value for the 
velocity and velocity dispersion of the object through a $\chi^2$-biased random 
walk. Then, a non-negative least squares algorithm \citep{lawson74} performs 
the inversion onto the stellar population model. Finally, the $\chi^2$-biased random 
walk is repeated to yield a best estimate for the kinematic parameters. 

Like the many other pixel-fitting softwares in existence 
\citep[e.g.][]{tojeiro07,ocvirk06,cid-fernandes05,koleva08}, 
this approach has the advantage of being able to fit the kinematics and composite 
stellar populations simultaneously. It also provides for a convenient way to mask 
spectral regions either in the observed frame (CCD defects, sky lines) and/or in the 
rest-frame (emission lines, regions badly fit by the models).


We caution that the fit results are sensitive to the continuum normalization 
method. Continuum-normalized spectral fitting has been favoured by e.g. 
\citet{wolf07} over other methods as the most reliable to simultaneously 
retrieve age and metallicity from the integrated spectra of stellar clusters.
After extensive testing, we settled for a simple running mean with 
exclusion of 3$\sigma$ outliers as the most robust procedure. 
The wavelength window over which the mean was evaluated was set to cover 
90 {\AA} \citep[compare][]{koleva08}.

\section{Application to Globular Clusters}
\label{s:fitres}


We now test the newly prepared differential models on high quality globular cluster 
spectra of NGC6528 and NGC6553 taken from \citet{schiavon05}  \footnote{The 
spectra can be downloaded from http://www.noao.edu/ggclib/description.html.}.
Only these two bulge GCs have [Fe/H] in the range probed by the differential models. 
For NGC6528 six spectra are available for consistency checks. 
Table \ref{t:sample} lists the published ages and abundances of these clusters. 
Not all literature 
measurements have assigned errorbars and we have been forced to assign 
hopefully realistic ones ad hoc.


\begin{table}
\begin{center}
\caption{Properties of GCs from colour-magnitude diagrams and high-resolution stellar spectroscopy}
\begin{tabular}{cccc}
Name    &   Age[Gyr]  &   [Fe/H]   & [$\alpha$/Fe] \\
\hline
NGC6528 & 11$\pm2$$^1$ & $-0.10\pm0.2$$^2$ & +0.1$\pm0.1$$^2$\\
NGC6553 & 11$\pm2$$^3$ & $-0.20\pm0.1$$^4$ & +0.25 $\pm0.1$$^3$\\
\end{tabular}
\label{t:sample}
\newline
$^1$ \citet{feltzing02}; $^2$ \citet{zoccali04}; $^3$ \citet{alves-brito06, melendez03}
\end{center}
\end{table}

%
%
%
%
%
%
%


We fit the spectra of the GCs using the three different differential model sets (BC03+C07, 
V09+C07, PegHR+C07). While an extensive test of the C07 models is beyond the scope
of this Letter, we also quote the results obtained with these models
alone to illustrate the improvement provided by the differential models.
The best-fitting SSP is determined by simply computing 
$\chi^2$ for every SSP with the three different parameters age, [Fe/H], [$\alpha$/Fe]. 
The only purpose of using {\tt sedfit} in this context is to automatically adapt for small 
shifts in velocity and for the resolution of the data. 

\begin{figure}[tbp]
\begin{center}
\resizebox{0.90\hsize}{!}{\includegraphics[]{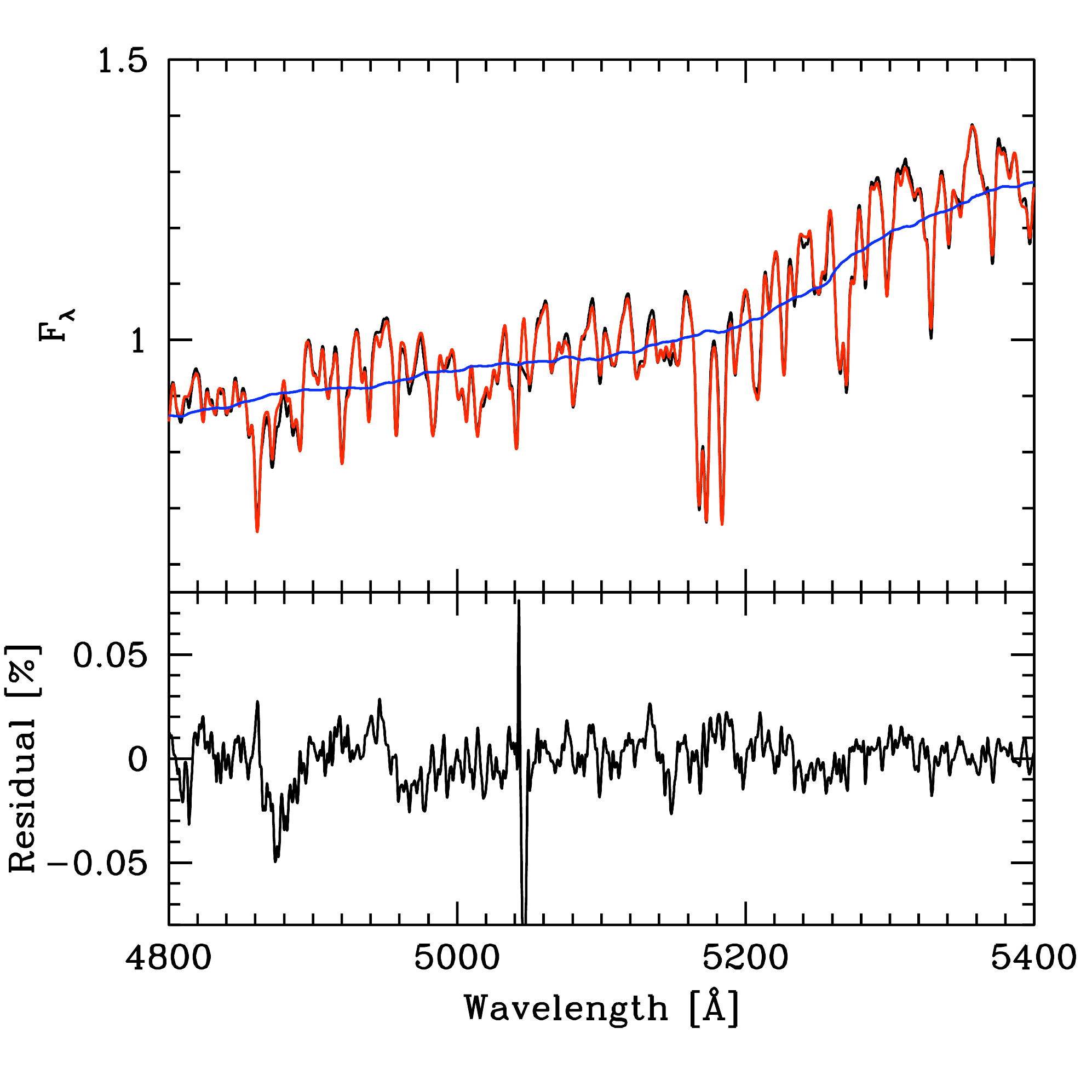}}
\caption[goodfit]{The spectrum of NGC6528 (black), labelled a\_2, as taken from \citet{schiavon05} 
and the best-fitting model from PegHR+C07 (red). The blue line is the continuum that was used to 
normalize the observed spectrum. The model was fit to the data only in the 
wavelength range 4828-5364 {\AA}. }
\label{f:goodfit}
\end{center}
\end{figure}

An example fit is shown in Figure \ref{f:goodfit}. Over the limited wavelength range 
of this fit, the model looks remarkable. A more quantitative measure is the reduced 
$\chi^2$, which is between 2 (for NGC6528\_b\_1 with PegHR+C07) and 10 (for 
NGC6528\_a\_1 with BC03+C07).  While thus the fit is not perfect, we note that an SSP with 
solar abundance ratios yields a reduced $\chi^2$ that is between 10\% and 50\% worse. 
Combining the realism of semi-empirical models and the predictive power of 
purely theoretical ones into the differential models thus leads to a significant improvement 
in the fit. 

We repeat the fits for different ranges in wavelength. From the outset, we 
exclude the wavelength region below 4000 {\AA}, as there are still not fully explained 
problems with semi-empirical models in this range \citep[e.g. ][]{wild07, walcher08}. 
We also exclude the region above 5800 {\AA}, which has some dubious features 
in the data. 

The wavelength regions we tested are, in order of increasing coverage: \\
R1: 5142-5364 {\AA} (MgB and Fe5335 indices) \\
R2: 4828-5364 {\AA} (main Lick index range)\\
R3: 4828-5600 {\AA} (red extension) \\
R4: 4500-5600 {\AA} (blue extension) \\
R5: 4000-5800 {\AA} (full wavelength range) \\

\begin{figure}
\begin{center}
\resizebox{0.90\hsize}{!}{\includegraphics[]{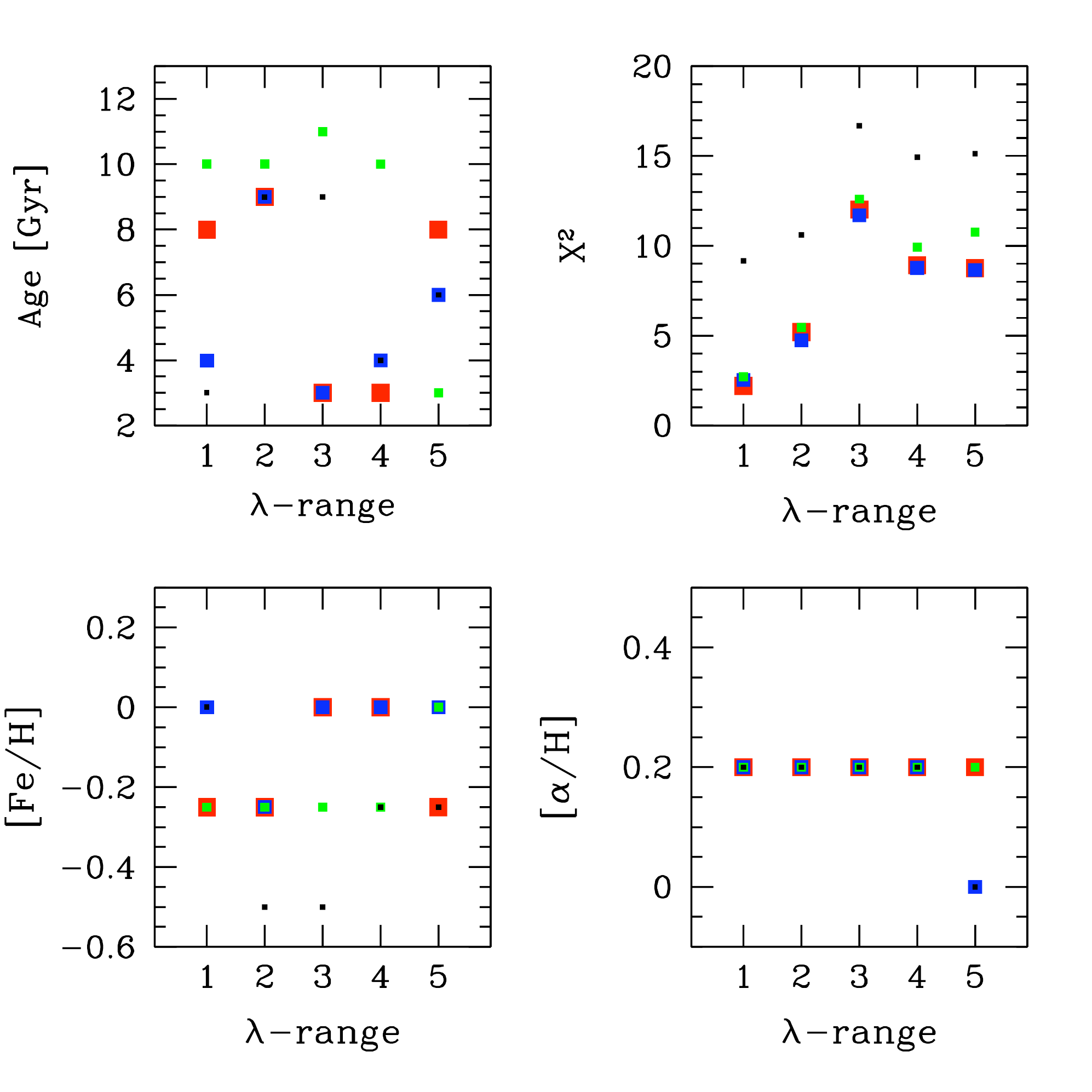}}
\caption[lamdelt]{The dependence of the best fit parameters on the wavelength range 
that was used for the fit and on the model. The numbers of the wavelength ranges are 
given in the text of Section \ref{s:fitres}. We show results for one spectrum 
(NGC6528\_a\_2) and for all models, i.e. V09+C07 
(red), PegHR+C07 (bue), BC03+C07 (green), C07 (black).}
\label{f:lamdelt}
\end{center}
\end{figure}

In Figure \ref{f:lamdelt} we show the dependence of the quartet reduced $\chi^2$, age, 
[Fe/H], and [$\alpha$/Fe] on the adopted wavelength range and for the spectrum 
NGC6528\_a\_2. Based on all plots for the 7 different GC spectra that we used, 
we find that for wavelength region R2 (4828 to 5364 {\AA}) the 
parameters [Fe/H] and [$\alpha$/Fe] 
of the best-fit SSP are in good agreement with the literature values for all differential models and 
all observed GC spectra. On the other hand, the results for the other wavelength ranges 
scatter more and seem in particular affected by the age-[Fe/H] degeneracy. 
In the case of R1, the available signal may not be strong 
enough. In the cases of large wavelength ranges, we expect that all models become 
less reliable, so small uncertainties in the modelling might dominate the fit results. 
Concerning age, we confirm the results of \citet{koleva08} that the spectral fitting ages for these 
clusters tend to be younger than the CMD ages. This is potentially attributable to contamination 
by blue bulge stars or blue straggler stars in the aperture covered by the integrated GC spectra 
and visible in the CMD (G. Beccari, priv. comm.) or to dispersion in the Helium abundances 
\citep{catelan09}, but we note also that the CMD ages do differ by a few Gyrs in the literature.

It seems to be of rather little relevance, which semi-empirical model one uses, as inside the 
regions in wavelength and parameter space that we test here, 
the differential models behave similarly. We nevertheless note that the PegHR+C07 
model yields the lowest overall $\chi^2$ but also systematically lower ages (and 
higher [Fe/H]), while fitting with the BC03+C07 models 
seems to be most stable with respect to wavelength range between 4500-5600 {\AA}. 
As expected, the fully theoretical models show a larger $\chi^2$ and the derived 
values tend to be less stable than those of the differential models. 

\begin{figure}
\begin{center}
\resizebox{0.90\hsize}{!}{\includegraphics[]{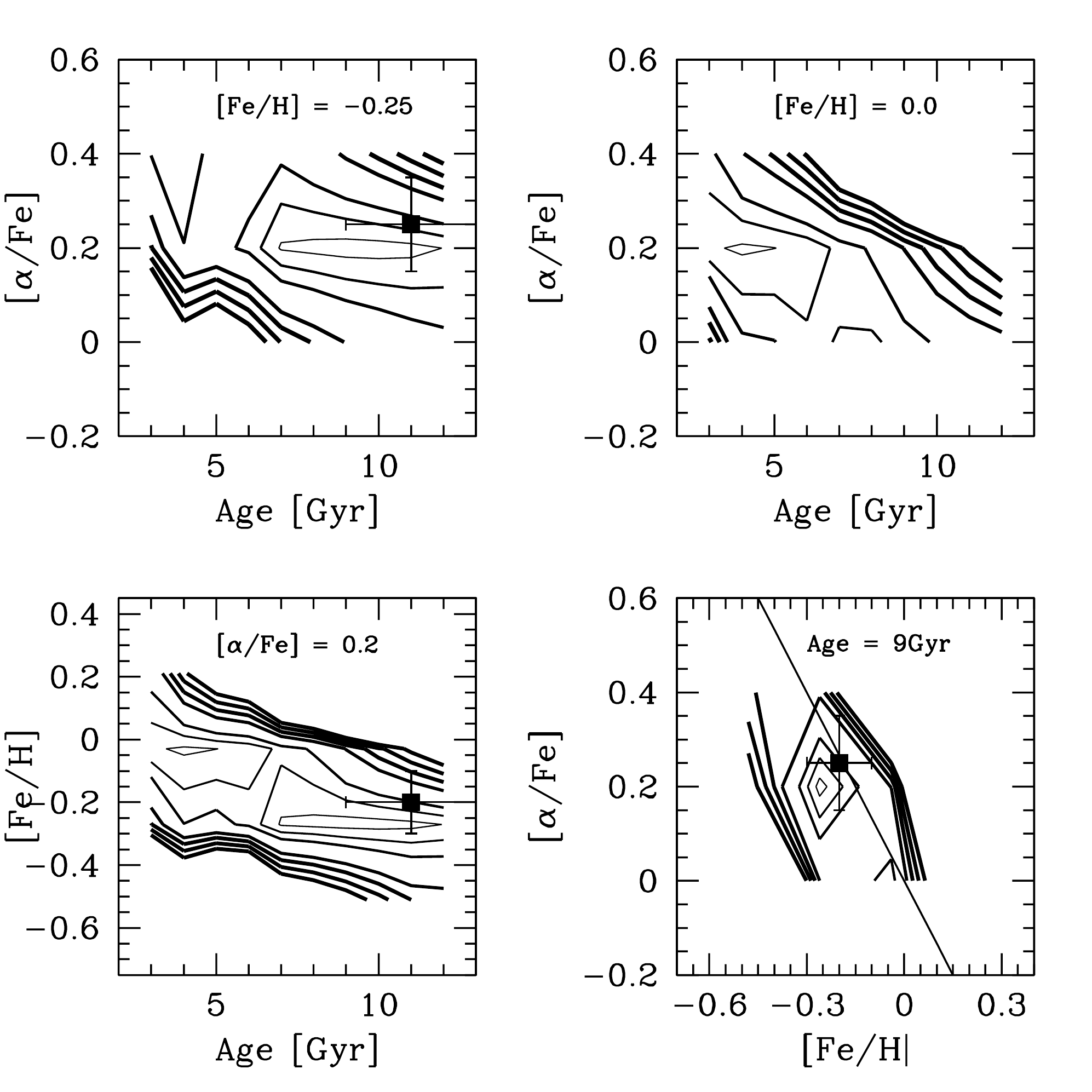}}
\caption[chi2plane]{The distribution of $\chi^2$ for the NGC6553\_a\_1 spectrum and determined 
over the wavelength region 4828-5364 {\AA}, depending on the SSP parameters age, [Fe/H], 
and [$\alpha$/Fe] using the V09+C07 model. The thicker the line the higher the value of $\chi^2$, 
in steps of $\Delta\chi^2=1$. The squares with errorbars show the literature values from 
Table \ref{t:sample}. The upper right panel shows that it is important to obtain both [Fe/H] 
and [$\alpha$/Fe] correctly, in order to avoid a biased age measurement. In the lower right 
panel the locus of constant total metallicity Z=0.017 is indicated by a thin straight line.} 
\label{f:chi2pl}
\end{center}
\end{figure}

In Figure \ref{f:chi2pl} we show the distribution of $\chi^2$ as a dependence of the three 
parameters age, [Fe/H], and [$\alpha$/Fe] for the NGC6553 spectrum.
The known age-[Fe/H] degeneracy is clearly visible. There is also a slight degeneracy 
between [Fe/H] and [$\alpha$/Fe],  following a line of constant total metallicity. 

\section{Test on SDSS early-type galaxies}

We now fit the spectra of a small sample of early-type galaxies from the SDSS survey 
\citep{abazajian08}. Early-type galaxies are known to be enhanced in $\alpha$-elements 
\citep{worthey92,trager00,vazdekis01,thomas04} and we expect this to be reflected in the best fit results. 
We select a subsample of 150 galaxies from the SDSS sample of \citet{gallazzi06}. 
The objects have been chosen in three bins of stellar mass, log(M$_*$/M$_{\odot}$) = 
[10-10.3], [10.7-11], [11.3-11.6]. In each mass bin we include only the first 50 
galaxies with the highest S/N and with $\Delta$(MgB/$<$Fe$>$) within 0.1 of the typical 
value for the mass range according to the linear fit
in \citet{gallazzi06} ($\Delta$(MgB/$<$Fe$>$) is an empirical estimate of [$\alpha$/Fe]). 

In order to test the validity of the differential models on the SDSS early-type spectra 
we use {\tt sedfit} to fit the star formation history of the galaxies in a non-parametric 
way with two sets of templates ranging in age from 3 to 12 Gyr: once with the 
templates [Fe/H] = -0.25 and [$\alpha$/Fe] = 0 and once with [Fe/H] = -0.25 and 
[$\alpha$/Fe] = 0.2. The fit was carried out using the V09+C07 model in the 
wavelength range 4828 to 5364 {\AA}. 

\begin{figure}
\begin{center}
\resizebox{0.90\hsize}{!}{\includegraphics[]{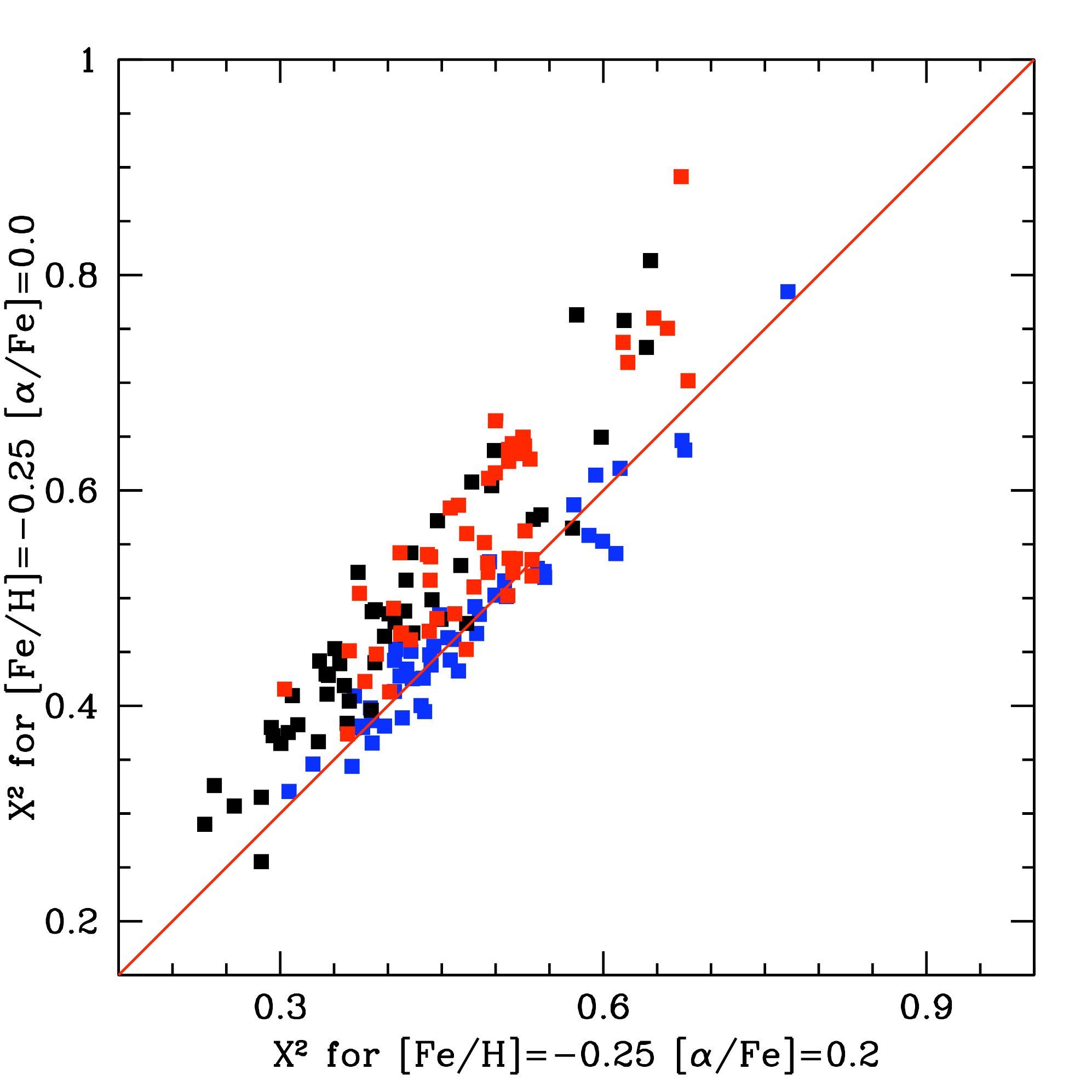}}
\caption[chi2plot]{The reduced $\chi^2$ of a composite age fit to the spectra of 
a sample of 150 early-type galaxies from the SDSS at fixed [Fe/H]=-0.25 and 
[$\alpha$/Fe]=0.2 (abscissa) and  [$\alpha$/Fe]=0.0 (ordinate).
Objects are split by mass, with the lowest in blue, intermediate in black 
and high-mass objects in red.}
\label{f:chi2plot}
\end{center}
\end{figure}

Figure \ref{f:chi2plot} shows that, while low-mass early-types are equally well fitted by 
scaled-solar than by alpha-enhanced models, for more massive early-types
[$\alpha$/Fe]=0.2 models provide on average a better fit. 
This reflects the expected mass-dependence of 
$\alpha$-enhancement in local early-type galaxies. We hold this result to show 
that the differential models represent the spectra of early-type galaxies in the SDSS
in a realistic way, over the whole spectrum between 4828 and 5364 {\AA}. 

The choice of values for [Fe/H] and [$\alpha$/Fe] in Figure \ref{f:chi2plot} is 
motivated by the distribution of the sample in these parameters as measured 
using the differential models. Nevertheless, due to the degeneracy between 
[Fe/H] and [$\alpha$/Fe] noted in Figure 4, similarly good fits are obtained 
when increasing [Fe/H] to 0.0, but    
the best fit [$\alpha$/Fe] would be shifted down by roughly 0.1 in compensation. 
A more thorough discussion of the impact of this degeneracy, 
a comparison with the global metallicities 
derived by \citet{gallazzi06} and the implications of the results
for the [Fe/H]  and [$\alpha$/Fe] distribution of SDSS early-type galaxies
will be presented in a forthcoming paper (Gallazzi et al., in prep.).

\section{Summary}

We have computed new, differential stellar population models. These combine 
(1) the realism of semi-empirical models at solar metallicities, due to 
the recently collected extensive libraries of stellar spectra, 
and (2) the predictive power of purely theoretical models, due to recent advances in the 
computation of purely theoretical stellar spectra \citep{percival09,coelho07}. 

We fitted the models to high-quality, public spectra of two globular clusters 
in the range of parameters covered by the new models. We showed that 
our best fit age, [Fe/H] and [$\alpha$/Fe] are in excellent agreement with those 
determined by other methods in the literature. Finally we fitted 
a sample of 150 early-type galaxies from the SDSS and show that the reduced $\chi^2$ 
is better for a model with [$\alpha$/Fe] = 0.2 than for one with [$\alpha$/Fe] = 0.0 for 
galaxies more massive than $\sim 10^{10.5}$M${\odot}$, again 
in accordance with independent measurements from indices.

The reduced $\chi^2$ values of the models on SDSS spectra are close to 1. We caution 
that this is not yet true for the highest S/N data of galactic globular clusters. Nevertheless, 
the uncertainties in the modelling are now \emph{smaller} than the effects of the varying 
abundances over a significant wavelength range (4828 to 5364 {\AA}). 
It needs to be pointed out though, that metallicity estimations from 
fitting the full optical spectrum, as done e.g. in \citet{panter08} and \citet{macarthur09}, 
are still rather uncertain. On the other hand, spectral fitting allows to use the redundant 
information on a large part of the spectrum and to adapt automatically for changing velocity 
dispersions, thus it yields tight constraints even for a large sample 
of intermediate S/N spectra such as those of the SDSS. 

We conclude that the differential models can reproduce accurately the detailed 
spectra of galaxies, thus allowing to measure [Fe/H] 
and [$\alpha$/Fe] in galaxies on an absolute scale and with high accuracy. 
With future improvements in semi-empirical and in theoretical models, differential 
stellar population models will become even more powerful.

\section*{Acknowledgments}

We thank the referee for valuable comments that led to an improved presentation 
of the results in this letter. 
P. Coelho is supported by a European Marie Curie incoming fellowship.

Funding for the SDSS and SDSS-II has been provided by the Alfred P. Sloan Foundation, 
the Participating Institutions, the National Science Foundation, the U.S. Department of 
Energy, the National Aeronautics and Space Administration, the Japanese Monbukagakusho, 
the Max Planck Society, and the Higher Education Funding Council for England. 
The SDSS Web Site is http://www.sdss.org/.
The SDSS is managed by the Astrophysical Research Consortium for the Participating Institutions. 


\bibliographystyle{mn2e}
\bibliography{biblio_difmod}

\end{document}